\documentclass[12pt]{article}
\usepackage{amsmath}
\usepackage{times}
\usepackage{graphicx}
\usepackage{color}
\usepackage{multirow}
\usepackage[authoryear]{natbib}
\usepackage{rotating}
\usepackage{bbm}
\usepackage{latexsym}

\newcommand{\captionfonts}{\normalsize}

\makeatletter  
\long\def\@makecaption#1#2{%
  \vskip\abovecaptionskip
  \sbox\@tempboxa{{\captionfonts #1: #2}}%
  \ifdim \wd\@tempboxa >\hsize
    {\captionfonts #1: #2\par}
  \else
    \hbox to\hsize{\hfil\box\@tempboxa\hfil}%
  \fi
  \vskip\belowcaptionskip}
\makeatother   

\begin{document}
\hspace{13.9cm}1

\ \vspace{20mm}\\

{\LARGE A Self-Organized Neural Comparator}

\ \\
{\bf \large Guillermo A. Ludue\~{n}a$^{\displaystyle 1}$ and Claudius
Gros$^{\displaystyle 1}$}\\
{$^{\displaystyle 1}$Institute for Theoretical Physics, Goethe University -  Frankfurt am Main, Germany.}\\

\ \\[-2mm]
{\bf Keywords:} Neural Networks, Self-organization, Complex Systems

\thispagestyle{empty}
\markboth{}{NC instructions}
\ \vspace{-0mm}\\
%
\begin{center} {\bf Abstract} \end{center}

Learning algorithms need generally the possibility 
to compare several streams of information. Neural learning
architectures hence need a unit, a comparator, able to 
compare several inputs encoding either internal or external
information, like for instance predictions and sensory readings. Without
the possibility of comparing the values of prediction to actual
sensory inputs, reward evaluation and supervised learning would
not be possible.
\newline
Comparators are usually not implemented explicitly, necessary
comparisons are commonly performed by directly comparing
one-to-one the respective activities. This implies that the
characteristics of the two input streams (like size and encoding) 
must be provided at the time of designing the system.
\newline 
It is however plausible that biological comparators emerge from
self-organizing, genetically encoded principles, which allow
the system to adapt to the changes in the input and in the organism.
We propose an unsupervised neural circuitry, where the function 
of input comparison emerges via self-organization only from the 
interaction of the system with the respective inputs, 
without external influence or supervision.
\newline 
The proposed neural comparator adapts, unsupervised,
according to the correlations present in the input streams.
The system consists of a multilayer feed-forward neural 
network which follows a local output minimization 
(anti-Hebbian) rule for adaptation of the synaptic weights. 
\newline
The local output minimization allows the circuit to autonomously 
acquire the capability of comparing the neural activities received 
from different neural populations, which may differ
in the size of the population and in the neural encoding used.
The comparator is able to compare objects never encountered 
before in the sensory input streams and to evaluate a measure 
of their similarity, even when differently encoded.

\section{Introduction}

In order to develop a complex targeted behavior, an autonomous 
agent must be able to relate and compare information received 
from the environment with internally generated information 
\citep[see][]{billing10}. For example it is often necessary 
to decide whether the visual image currently being perceived 
is similar to an image encoded in some form in memory. 

For artificial agents, such basic comparison capabilities are
typically either hard-coded or initially taught, both processes
involving the inclusion of predefined knowledge \citep[for
instance][]{bovet05a,bovet05b}. However, living organisms must acquire
this capability autonomously, only via interaction with the
acquired data, possibly without any explicit feedback from the
environment \citep{oreilly-munakata}.  We can therefore hypothesize
the presence of a neural circuitry in living organisms, capable of
comparing the information received by different populations of
neurons. It cannot therefore in general be assumed that these
populations have a similar configuration, hold the information in the
same encoding or even manage the same type of information.

A system encompassing said characteristics must be based on some form
of unsupervised learning, it must self-organize in order to
autonomously acquire its basic functionality. The task of an
unsupervised learning system is to elucidate the structure in the
input data, without using external feedback. Thus all the
information should be inferred from the correlations found in the
input and in its own response to the input data stream.

Unsupervised learning can be achieved using neural networks, and has
been implemented previously for a range of applications \citep[see for
instance][]{sanger-optimal,Atiya1990,Likhovidov1997,Furao2007,Tong2008}.
Higher accuracy is generally expected from supervised algorithms.
However, \cite{Japkowicz95anovelty,japkowicz-binary} have shown that
for the problem of binary classification, unsupervised learning in a
neural network can perform better than standard supervised approaches
in certain domains.

Neural and non-biologically inspired algorithms are often expressed in mathematical
terms based on vectors and their respective elementary operations like
vector subtraction, conjunction and disjunction. Implementations in
terms of artificial neural networks hence typically involve the
application of these operations to the output of groups of neurons.
These basic operations are however not directly available in
biological neural circuitry, which is based exclusively on local
operations.  Connections between groups of neurons evolve during the
growth of the biological agent and may induce the formation of
topological maps \citep[see][]{kohonen1990self} but generically do not
result in one-to-one neural operations. For instance these one-to-one
neural interactions would involve an additional global summation of
the result for the case of a scalar product.
 
It is unclear whether operations like vector operations are directly
used by biological systems, their implementation should be in any case
robust to the changes in the development of the system and to its
adaptation to different types of input.  In effect, the basic
building blocks of most known learning algorithms are the mathematical
functions computers are based on. These are, however, not necessarily
present, convenient or viable in a biological system.  It is our aim
to elucidate how a basic mathematical function can emerge naturally in
a biological system.  We present, for this purpose, a model of how the
basic function of comparison can emerge in an unsupervised neural
network, based on local rules for adaption and learning.  Our
adaptive ``comparator'' neural circuit is capable of self-organized
adaption, with the correlations present in the data input stream as
the only basis for inference.
 
\begin{figure}[t]
\centering
\includegraphics[width=0.6\textwidth]{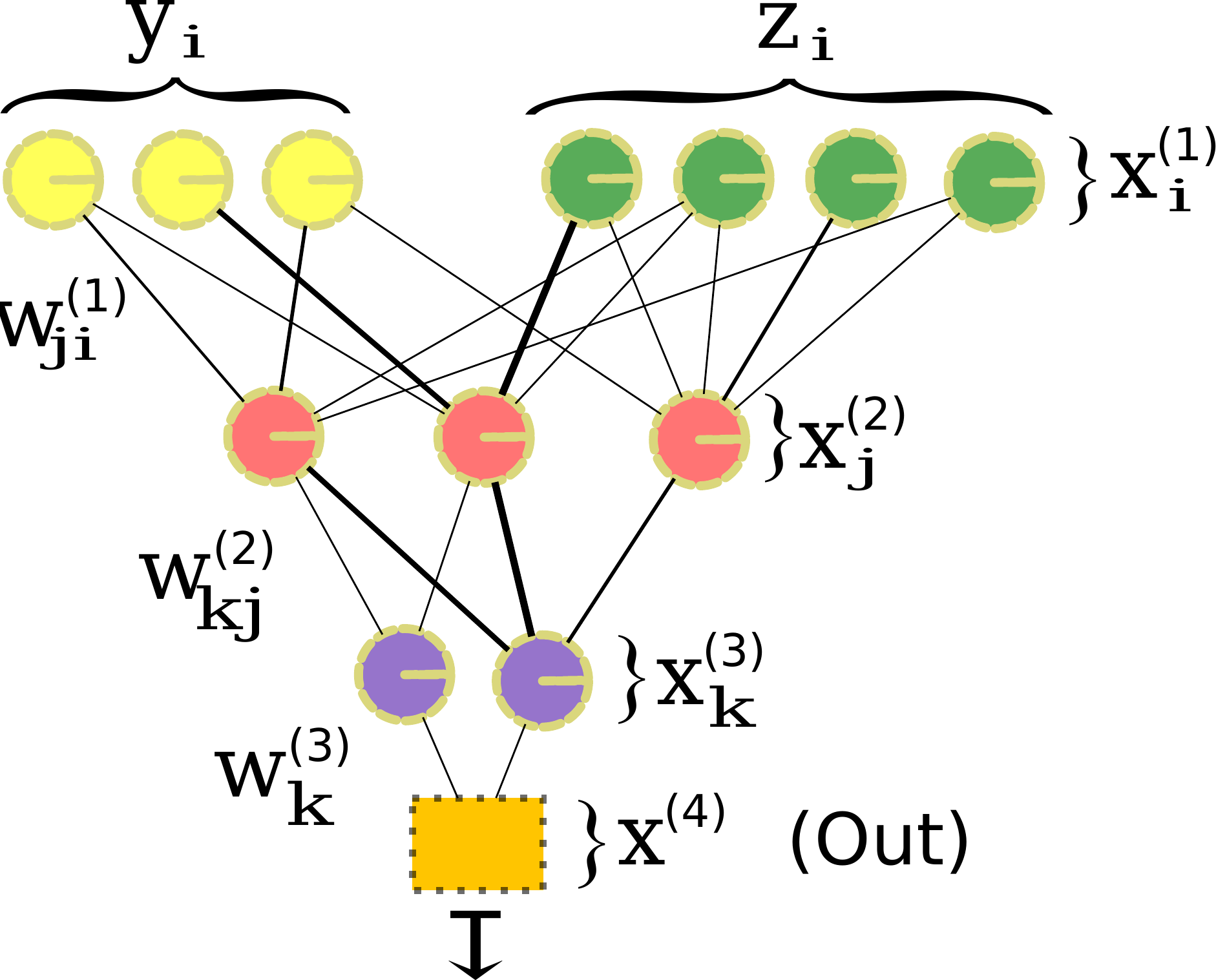}
\caption{Architecture of the proposed neural comparator, with
         intermediate layers $\mathbf{x}^{(2)}$ and 
         $\mathbf{x}^{(3)}$ and output $\mathbf{x}^{(4)}$. 
         The comparator autonomously learns to compare the inputs 
         $\mathbf{y}$ and $\mathbf{z}$. The output is
         close to zero when the two inputs are identical or
         correlated and close to unity when they are different
         or uncorrelated.
         }
\label{fig:scheme}
\end{figure}

The circuit autonomously acquires the capability of comparing the
information received from different neural populations, which may
differ in size and in the encoding used.  The comparator proposed  is
based on a multilayer feed-forward neural network, where the input
layer receives two signals $\mathbf{y}$ and $\mathbf{z}$, see
fig.~\ref{fig:scheme}. These two input streams can be either
unrelated, selected randomly or, with a probability, encode the
same information. The task of the neural comparator is then to
determine, for any pair of input signals $\mathbf{y}$ and
$\mathbf{z}$, whether they are semantically related or not.
Any given pair ($\mathbf{y}$,$\mathbf{z}$) of semantically
related inputs is presented to the system, in general, only one
single time. The system has hence to master the task of
discriminating generically between related and unrelated pairs 
of inputs, and not the task to extract statistically
repeatedly occurring patterns.

The strength of the synapses connecting neurons are readjusted using
anti-Hebbian rules. Due to the readjustment of the synaptic weights,
the network minimizes its output without the help of external
supervision. As a consequence, the network is able to autonomously
learn to discriminate whether the two inputs encode the same
information or not, independently of whether the particular input
configuration has been encountered before or not. The system will
respond with a large output activity whenever the input pair
($\mathbf{y}$,$\mathbf{z}$) is semantically unrelated and with
inactivity for related pairs.

\subsubsection{Motivation and expected use case}

We are motivated by a system where the information stored in two
different neuronal populations are to be compared. In particular, we
are interested in systems as the one presented by \cite{bovet05a},
where two streams of information (for instance, visual input and the
desired state of the visual input, or the signal from the whiskers of
a robot compared to the time-delayed state of these sensors) encoded in
two separate neuronal populations are to be compared, in this
particular case in order to get a distance vector between the two.  In
a fixed artificial system, one could obtain this difference by simply
subtracting the input from each of the streams, provided that the two
neuronal populations are equal and encode the information in the same
way. This subtraction can also be implemented in such a system in a
neuromorphic way simply by implementing a distance function in a
neural network. However, we are interested in the case where both
neuron populations have evolved mostly independently, such that they
might be structurally different and might encode the information in a
different way, which is expected in a biological system. Under these
conditions, the neuronal circuit comparing both streams should be able
to invert the encoding of both inputs in order to compare them, a
task which could not be solved using a fixed distance function.
In addition, we expect that such a system would be deployed in an
environment where it is more probable to have different, semantically
unrelated, inputs than otherwise. The comparator should hence
be able to solve the demanding task of autonomously extracting
semantically related pairs of inputs out of a majority 
of unrelated and random input patterns.

\section{Architecture of the neural comparator}

The neural comparator proposed consists of a feed-forward 
network of three layers, plus an extra layer filtering the 
maximum output from the third layer, compare fig.~\ref{fig:scheme}.
We will refer to the layers as $k=1,2,3,4$, where 
$k=1$ corresponds to the input layer and $k=4$ 
to the output layer. The output of the individual neurons is 
denoted by $x_i^{(k)}$, where the supraindex refers to the layer 
and the subindex to the index of the neuron in the layer, 
for instance $x_2^{(1)}$ being the output of the second neuron in the
input layer.

The individual layers are connected via synaptic weights 
$w_{ji}^{(k)}$. In this notation, the index $i$ 
corresponds to the index of the presynaptic neuron in layer $k$, 
and $j$ corresponds to the index of the postsynaptic neuron in 
layer $k+1$. Thus $w_{3,4}^{(1)}$ is the synaptic weight connecting 
the fourth input neuron with the third neuron in the second layer.

The layers are generally not fully interconnected.
The probability of a connection between a neuron in layer $k$
and a neuron in layer $k+1$ is $p_{conn}^{(k)}$. The values
used for the interconnection probabilities $p_{conn}^{(k)}$ 
are given in table~\ref{tab:params}.

In the implementation proposed and discussed here, the output layer is
special in that it consists only in selecting the maximum of all
activities from the third layer. There are simple neural architectures
based on local operations that could fulfill this purpose. However,
for simplicity, the task of selecting the maximum activity of the third
layer is done here directly by a single unit.

\subsection{Input protocol}
\label{sec:protocol}

The input vectors $\mathbf{x}^{(1)}$ consist of two parts
\begin{equation}
\mathbf{x}^{(1)} \ =\ \big(\mathbf{y}, \mathbf{z}\big)~,
\label{eq:x_1}
\end{equation}
where $\mathbf{y}$ and $\mathbf{z}$ are the two 
distinct input streams to be compared. We used
the following  protocol for selecting pairs of
$(\mathbf{y},\mathbf{z})$:
\begin{itemize}
\item $\mathbf{y}$ is selected randomly at
      each time step, with the elements $y_i \in [0,1]$ 
      drawn from a uniform distribution. 
\item $\mathbf{z}$ is selected via
\begin{equation}
\mathbf{z} \ =\ \left\{
\begin{array}{rl}
\mathrm{random} & \mathrm{with\ probability}\ 1-p_{eq}\\
\mathbf{f}(\mathbf{y}) & \mathrm{with\ probability}\ p_{eq}
\end{array}
\right.
\label{eq:z}
\end{equation}
\end{itemize}
If the inputs $\mathbf{z}$ and $\mathbf{y}$
carry the same information, they are related
via $\mathbf{z} =\mathbf{f}(\mathbf{y})$,
where $\mathbf{f}$ is generically an injective transformation.
This relation reduces to $\mathbf{z}=\mathbf{y}$ for the case
when the encodings in the two neural populations 
$y$ and $z$ are the identity. 

We consider two kinds of encoding, direct encoding with
$\mathbf{f}$ being the identity, and encoding through a
li\-near transformation, which we refer as ``linear encoding'',
\begin{equation}
\mathbf{z}= \mathbf{y} \qquad \mathrm{or}\qquad 
\mathbf{z}=\hat A \mathbf{y} \; , 
\end{equation}
where $\hat A$ is a random matrix. The encoding is maintained
throughout individual simulations of the comparator.  For the case of
linear encoding, the matrix $\hat A$ is selected initially and not
modified during a single run.
 
The procedure we used to generate the matrix $\hat A$ consists of
choosing each element of the matrix as a random number taken from a
continuous flat distribution of values between -1 and 1. The matrix is
then normalized such that the elements of vector $\mathbf{z}$ belong
to $z_i=[-1,1]$.

\subsection{Synaptic weights readjustment: anti-Hebbian rule}

Each neuron integrates its inputs, via
\begin{equation}
x_j^{(k+1)}=g\left(\sum_i{w^{(k)}_{ji} x_i^{(k)}}\right),
\qquad
g(x)=\tanh(\alpha x) ~,
\label{eq_update_rule}
\end{equation}
with $g(x)$ being the transfer function, $\alpha$ the
gain and $w^{(k)}_{ji}$ the afferent synaptic weights.
After the information is passed forward, the synaptic 
weights are corrected using an anti-Hebbian rule,

\begin{eqnarray}
\nonumber
\Delta w_{ji}^{(k)}(t) &\equiv& w_{ji}^{(k)}(t+1) - w_{ji}^{(k)}(t) \\
&=& - \eta x_i^{(k)}(t) x_j^{(k+1)}(t) \; .
\label{eq:updatew}
\end{eqnarray}
Neurons under an anti-Hebbian learning rule will modify 
their synaptic weights in order to minimize their output.
Note, that anti-Hebbian adaption rules generically result 
from information maximization principles \citep[][]{bell95}.
Information maximization favors spread-out output activities 
for statistically independent inputs \citep[][]{Markovic2010}, 
allowing such to filter-out correlated input pairs 
$(\mathbf{y}, \mathbf{z})$, which tend to induce a low level of
output activities.

The incoming synaptic weights to neuron $i$ of the $(k+1)$th 
layer are additionally normalized, after an update of 
the synaptic weights, 
\begin{equation}
w_{ji}^{(k)}(t+1)\ \to\ 
\frac{w_{ji}^{(k)}(t+1)}{\sqrt{\sum_j \big| 
w_{ji}^{(k)}(t+1) \big|^2}} \; .
\label{eq:renormw}
\end{equation}

The algorithm proposed here is based on the idea that 
correlated inputs will lead to a small output, as a 
consequence of the anti-Hebbian adaption rule.
Uncorrelated pairs of input $(\mathbf{y},\mathbf{z})$
will on the other side generate, in general, a substantial 
output, as they correspond to random inputs for which the
synaptic weights are not adapted to minimize the output.
It is worthwhile to remark that using a Hebbian 
adaption rule and classifying the minimum values 
as uncorrelated would not achieve the same accuracy 
as with the proposed anti-Hebbian rule with output
values between -1 and 1. The reason is that we seek
a comparator capable of comparing arbitrary pairs
$(\mathbf{y},\mathbf{z})$ of input, and not 
specific examples.

When using an anti-Hebbian rule, zero output is an 
optimum for any correlated input. In the case of input 
with equal encoding, this is reached when the synaptic 
weights cancel exactly ($w_{left}=-w_{right}$) in the first
layer, compare fig.~\ref{fig:scheme}. In contrast, if 
a Hebbian rule would be used, the optimum value for 
correlated input corresponds to the synaptic weights 
of correlated input being as large as possible. 
The consequence is that all synaptic weights tend 
to increase constantly, eventually leading to all 
output achieving maximum values.

There remains, for anti-Hebbian adaption rules, a 
statistically finite probability for uncorrelated 
inputs to have a low output by mere chance, viz
the terms $w_{ji}^{(k)}x^{(k)}_i$ originating from
$\mathbf{y}$ and $\mathbf{z}$ may cancel out. In 
such cases the comparator would be misclassifying 
the input. The occurrence of misclassification 
is reduced substantially by having multiple neurons 
in the third layer.

By selecting an inter-layer connection probability
$p_{conn}^{(2)}$ well below unity, the individual
neurons in the third layer will have access to
different components of the information encoded in
the second layer. This setup is effectively equivalent 
to generating different and independent parallel paths 
for the information transfer, adding robustness to the 
learning process, since only strong correlations 
between the input pairs
$(\mathbf{y},\mathbf{z})$, shared by the majority of
paths, are then acquired by all neurons.

In addition to diminishing the possibility of random
misclassification due to the multiple paths, the use of anti-Hebbian
learning in the third layer minimizes the incidence of the individual
parallel paths which consistently result in $x_i^{(2)}$ outputs that
are far larger than the rest (\emph{failing} paths, since they are
unable to learn some correlations). Thus adding this layer results in
an significant increase in accuracy with respect to an individual
2-layer comparator.  The accuracy is further improved by adding a
filtering layer for input classification.

\subsection{Input classification}

Each third-layer neuron encodes the estimated degree of correlation
within the input pairs, $(\mathbf{y},\mathbf{z})$. The fourth
layer selects the most active third-layer neuron, 

\begin{equation}
x^{(4)}\ =\ \max_j |x_j^{(3)}| \; .
\label{eq:max}
\end{equation}

By selecting the maximum of all outputs in the third layer, the circuit
looks for a ``consensus'' among the neurons in the third layer. A
given input pair ($\mathbf{y}$,$\mathbf{z}$) needs to be considered as
correlated by all third-layer neurons in order to be classified as
correlated by the fourth layer. This, together with the randomness of
the inter-layer connections, increases the robustness of the
classification process.

There are several options for evaluating the effectiveness of
the neural comparator. We will later discuss,
in sect.~\ref{sec:fuzzy}, an analysis in terms of fuzzy logic,
and consider now a measure of the accuracy of the system
in terms of binary classification.
The inputs $\mathbf{y}$ and $\mathbf{z}$ are classified according to
the strength of the value of $x^{(4)}$. For binary classification
we use a simple threshold criterion. The inputs 
$\mathbf{y}$ and $\mathbf{z}$ are considered to be 
uncorrelated if
$$ x^{(4)}\ >\ \theta \; , $$
and otherwise correlated. In this work, the value for the threshold
$\theta$ is determined by minimizing the probability of
misclassification, in order to test the possible accuracy of the
system. The same effect of this minimization could be achieved by
keeping $\theta$ fixed and optimizing the slope $\alpha$ of the 
transfer function (\ref{eq_update_rule}), since $\theta$ depends 
on the slope $\alpha$.
These parameters, the slope $\alpha$ or the discrimination
threshold $\theta$, may in principle be optimized autonomously
using information theoretical objective functions
\cite[][]{Triesch2005,Markovic2010}. For simplicity we here perform
the optimization directly. We will show in sec.~\ref{sec:impactpeqandn}, 
that the optimal values for $\alpha$ and $\theta$ depend
essentially only on the size $N$ of the input. Minor adjustments 
of the parameters might anyway be desirable to maintain an optimal 
accuracy. In any case, these readjustments can be done in a biological 
system via intrinsic plasticity 
\citep[see][]{stemmler1999,Mozzachiodi2010,Markovic2011}.

Although we did not implement the $max$ function present in
(\ref{eq:max}) in a neuromorphic form, a small neuronal circuit
implementing (\ref{eq:max}) could for instance be realized as a
winner-takes-all network \citep{Coultrip1992,kaski1994,Grossberg1987}.
Alternatively, a filtering rule different from the $max$ function
could be used for the last layer, for instance the addition or
averaging of all the inputs.
We present as supporting information some results showing the behavior
of the output when using averaging and sum as alternative filtering
rules for the output layer.
Our best results were however found by implementing the last layer as
a $max$ function. In this work we will discuss the behavior of the
system using the $max$ function as the last layer.

We would like to remark that defining a threshold $\theta$
is one way of using this system for binary classification, which we
use for reporting the possible accuracy of the system. However, it is
not a defining part of the model. We expect the system to be more
useful for obtaining a continuous variable measuring the grade of
correlation of the inputs. As we discuss in sec.~\ref{sec:fuzzy}, this
property can be used to apply fuzzy logic in a biological system.

\section{Performance in terms of binary classification}

\subsection{Performance measures}

In order to access the performance, in terms of binary classification,
of the neural comparator, we need to track the numbers of correct and
incorrect classifications. We use the following three measures for
classification errors:
\begin{itemize}

\item[$FP$] \underline{false positives}:
    The fraction of cases for which
    $x^{(4)}<\theta$ (input is classified as correlated) 
    occurs for uncorrelated pairs of
    input vectors $\mathbf{y}$ and $\mathbf{z}$:
\begin{equation}
FP\ =\ \frac{\mathrm{\#\,erroneously\, assigned\, as\, positive}}
            {\mathrm{\#assigned\, as\, positive}} \; .
\label{eq:FPdef}
\end{equation}
\item[$FN$] \underline{false negatives}:
    The fraction of cases with output activity $x^{(4)}\ge\theta$ 
    (input classified as uncorrelated) occurring for
    correlated pairs of input vectors,
    $\mathbf{z}=\mathbf{f}(\mathbf{y})$:
\begin{equation}
FN\ =\ \frac{\mathrm{\#\,erroneously\, assigned\, as\, negative}}
            {\mathrm{\#assigned\, as\, negative}} \; .
\label{eq:FNdef}
\end{equation}
\item[$E$] \underline{overall error}:
           The total fraction of errors $E$ is the 
           fraction of overall wrong classifications:
\begin{equation}
E\ =\ \frac{\mathrm{\#\,erroneously\, assigned}}
           {\mathrm{\#all\, assignments}} \; .
\label{eq:Edef}
\end{equation}
\end{itemize}
All three performance measures, $E$, $FP$ and $FN$, need to be kept
low.  This is achieved for a classification threshold $\theta$ which
minimizes $(FP+FN)$. This condition keeps all three error measures,
$FP$, $FN$ and $E$ close to their minimum, while giving $FN$ and $FP$
equal importance at the same time.

\begin{figure}[!ht]
\centering
\includegraphics[width=0.7\textwidth]{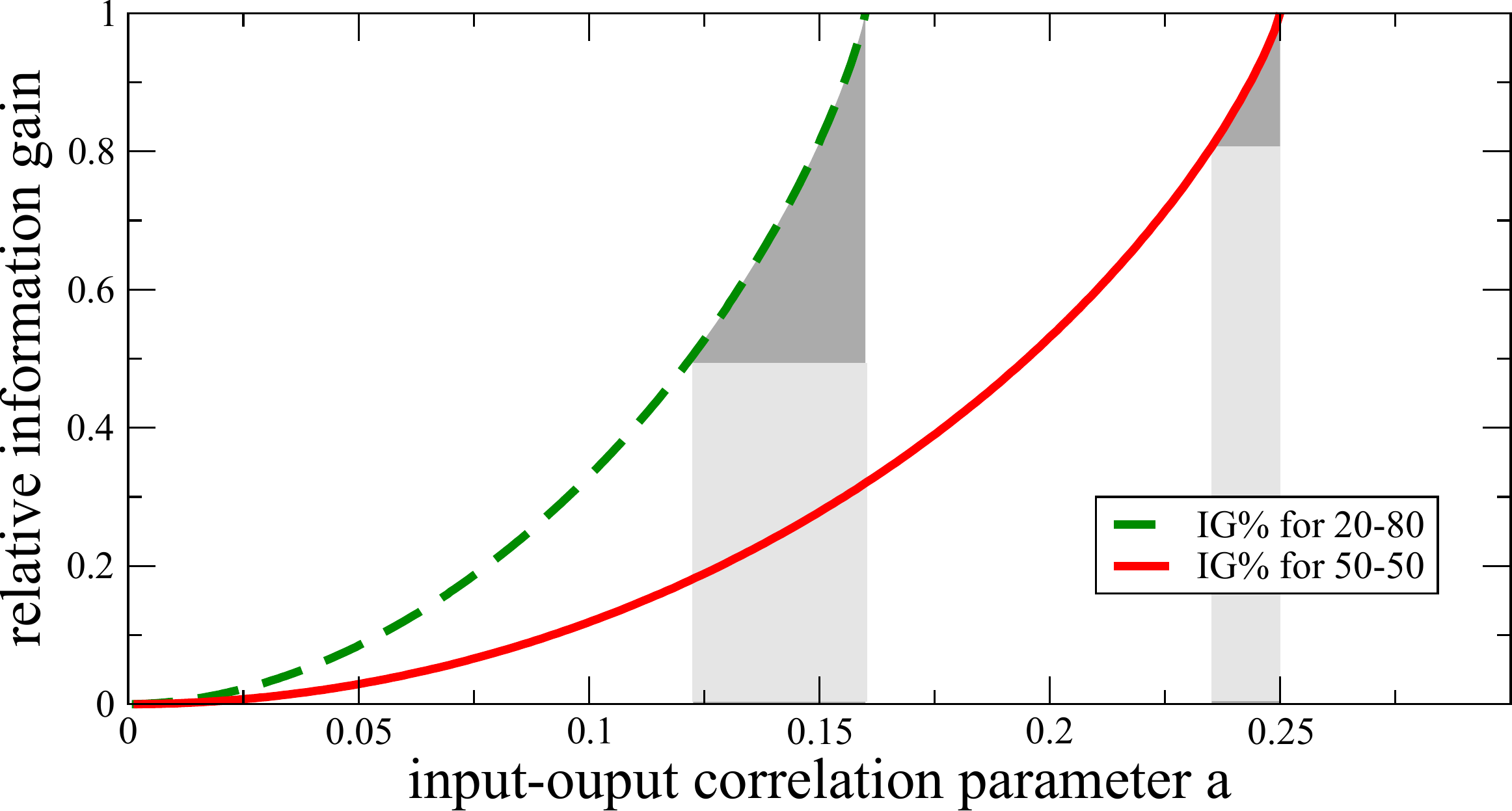}
\caption{The relative mutual information MI\% (\ref{eq_MI_relative}),
as a function of the amount of correlation $a$ between the
input and the output, defined by eq.~\ref{eq:MI_model}.
The comparator achieves an MI\% of about 50\% and 80\%
respectively for fractions of $p_{eq}=0.2$ (dashed line)
and $p_{eq}=0.8$ (solid line) of correlated input pairs,
corresponding to $0.12/0.16=0.75$ and $0.23/0.25=0.92$
of all input-output correlations $a$. Hence only 
25\% and 8\%, respectively
(darker shaded areas), of all correlations
are not recovered by the comparator.
}
\label{fig:MI_relative}
\end{figure}

\subsubsection{Mutual Information}

Since the percentage of erroneous classifications, despite it being an
intuitive measure, is dependent on the relative number of correlated
and uncorrelated inputs presented to the system, we also evaluate the
mutual information (MI) \citep[see for instance][]{gros2008} as a
measure of the information that has been gained by the classification
made by the comparator:
\begin{eqnarray}
\label{eq_MI}
    \mathrm{MI}(X,Y)  & = & H[X] - H[X|Y] \\
    & = & - \sum_{x \in X} p(x) \log{(p(x))} +
    \sum_{y\in Y} p(y) \sum_{x \in X} p(x|y) \log{(p(x|y))} \; ,
\nonumber
\end{eqnarray}
where $X$ in this case represents whether the inputs are equal or not,
and $Y$ is whether the comparator classified the input as correlated
or not, therefore, both $X$ and $Y$ are vectors of size two 
($true/false$ corresponding to semantically related/uncorrelated). 
Here $\rho(x|y)$ is the conditional probability that the input had 
been $x=frue/false$ given that output of the comparator is 
$y=true/false$ and $H(X)$ the marginal information entropy.

We will refer specifically to the mutual information (\ref{eq_MI})
between the binary input and output of the neural comparator, also
known in this context as information gain. The mutual information
(\ref{eq_MI}) can also be written as
$\sum_{x,y}p(x,y)\log(p(x,y)/(p(x)p(y)))$, where $p(x,y)=p(y|y)p(y)$
is the joint probability. It is symmetric in its arguments $X$ and $Y$
and positive definite. It vanishes for uncorrelated processes $X$ and
$Y$, viz when $\rho(x|y)=\rho(x)$, viz for a random output of the
comparator. Finally, the mutual information is maximal when the two
processes are 100\% correlated, that is, when the off-diagonal
probability vanish, $\rho(x|y)=0$ for $x\ne y$.  In this case the
two marginal distributions $\rho(x)=\sum_y p(x,y)$ and $\rho(y)=\sum_x
p(x,y)$ coincide and $MI(X,Y)$ is identical to coinciding marginal
entropies, $ H(X)=H(Y) $.

We will present the mutual information as the percentage of the maximally
achievable mutual information,
\begin{eqnarray}
\label{eq_MI_relative}
    \mathrm{MI\%}(X,Y)  & = & \frac{H[X] - H[X|Y]}{H[X]} \\
    & = & 1 - 
    \frac{\sum_{y\in Y} p(y) \sum_{x \in X} p(x|y) \log{(p(x|y))}}{\sum_{x \in X} p(x) \log{(p(x))}} \; ,
\nonumber
\end{eqnarray}
which has hence a value between 0 and 1, and is therefore more intuitive
to read as a percentage of the maximum theoretically possible. The
maximum mutual information achievable by the system depends solely on
the probabilities of correlation/decorrelation, i.e. $p_{eq}$.

The statistical correlations between the input $X$ and 
the output $Y$ can be parametri\-zed using a correlation
parameter $a$, via
\begin{equation}
	p(x,y) \ =\ p(x)\,p(y) \ + C_{a}(x,y)
\label{eq:MI_model}~,
\end{equation}
where $C_{a}(x,y)$ are the element values of the matrix:
\begin{equation}
	C_{a} =
	\left(\begin{array}{cc}
	a & -a \\ -a & a
	\end{array}\right)
	\qquad\quad
	a\in\big[0,p_{eq}(1-p_{eq})\big]
\label{eq:MI_model_mat}~.
\end{equation}
Here $p_{eq}$ is the probability of having correlated
pairs of inputs, viz $p(x=true)=p_{eq}$
and $p(x=false)=(1-p_{eq})$.
Using this parametrization allows us to evaluate
the relative mutual information (\ref{eq_MI_relative})
generically for a correlated joint probability
$p(x,y)$, as illustrated in fig.~\ref{fig:MI_relative}.
The parametrization (\ref{eq:MI_model}) hence provides
an absolute yardstick for the performance of the
neural comparator.

\subsection{Simulation results}

We performed a series of simulations using the network parameters
listed in table~\ref{tab:params}, and for two encoding rules
(\ref{eq:z}), direct and linear encoding. The lengths $N$ of the input
vectors $\mathbf{y}$ and $\mathbf{z}$ are taken to be equal, if not
stated otherwise.

\begin{table}[bt]
\centering
\begin{tabular}{c||c|c|c|c|c}
  Param.  & $t_{max}$ & $N^{(1)}$ & $N^{(2)}$ & $N^{(3)}$ &  $\alpha$ ($N< 400$)  \\\hline
  Value   & $10^7$ & 2$N$ & $N$ & $\lfloor \frac{N+1}{2}\rfloor$ & 2.7 \\\hline\hline
  Param.  & $p_{conn}^{(1)}$ & $p_{conn}^{(2)}$ & $\eta$ & $p_{eq}$ & $\alpha$ ($N\ge 400$)\\\hline
  Value   & 0.8 & 0.3 & 0.003 & 0.2 & 1.0 \\
\end{tabular}
\caption{Parameters used in the simulations. 
         $N$: Input vector size, $N^{(k)}$: size of layer $k$, 
         $t_{max}$: number of steps of simulation, 
         $p_{eq}$: probability of equal inputs, 
         $p_{conn}^{(k)}$: probability of connection from 
         the $k$th layer to the $(k+1)$th layer, 
         $\alpha$: sigmoid slope, $\eta$: learning rate.}
\label{tab:params}
\end{table}

\subsubsection{Low Probability of Equals}

Since our initial motivation for the design of this system is the
comparison of two input streams that are presumably most of the time
different, we have studied the behavior of the system when there is a
lower probability of an event where both streams are equal than
otherwise.  We used $p_{eq}=0.2$ in (\ref{eq:z}), viz in 20\% of the
cases the relation $\mathbf{z}=\mathbf{f}(\mathbf{y})$ holds and in
the remaining 80\% the two inputs $\mathbf{y}$ and $\mathbf{z}$ are
completely uncorrelated (randomly drawn). Each calculation consists of
$t_{max}=10^7$ steps, from which the last 10\% of the simulation is
used for the evaluation of the performance.  During said last 10\% of
the simulation the system keeps learning, i.e. there is no separation
between training and production stages. The purpose of taking only
the last portion is to ignore the initial phase of the learning
process, since at that stage the output does not provide a good
representation of the system's accuracy.

In table~\ref{tab:errs}, we present the mean values for the different
measures of error, eqs.~(\ref{eq:FPdef})--(\ref{eq:Edef}), observed
for 100 independent simulations of the system.  For each individual
simulation, the interlayer connections are randomly drawn with
probabilities $p_{conn}^{(k)}$, the parameters are as shown in table
\ref{tab:params}. The errors for each run are calculated using a
threshold $\theta$ that minimizes the sum of errors $\langle
FP\rangle+\langle FN\rangle$. Each input in the first layer has a
uniform distribution of values between -1 and 1. The accuracy of the
comparator is generally above 90\%, in terms of binary classification
errors. There is, importantly, no appreciable 
difference in the accuracy when using direct encoding or
linear encoding with random matrices.

Note, that a relative mutual information of 
MI\%$\approx$50\% is substantial \citep[][]{Guo2005}.
A relative mutual information of 50\% means that the
correlation between the input and the output of
the neural comparator encompasses 75\% of the 
maximally achievable correlations, as
illustrated in fig.~\ref{fig:MI_relative}.

\begin{table}[t]\centering
\begin{tabular}{c||c|c|c|c|c|c|c|c}
& \multicolumn{4}{|c}{direct} & \multicolumn{4}{|c}{linear} \\
\hline
$N$ & $\langle E \rangle$ & $\langle FP \rangle$ & $\langle FN
\rangle$ & MI\% & $\langle E \rangle$ & $\langle FP
\rangle$ & $\langle FN \rangle$ & MI\% \\\hline\hline
5   & 10.2\%  &  5.8\% & 10.5\% & 13.2\%& 14.8\% & 8.3\% & 14.8\% & 23.8\% \\
15  &  6.0\%  &  1.2\% &  6.8\% & 44.4\%&  5.2\% & 2.8\% &  5.9\% & 41.7\% \\
30  &  5.3\%  &  1.0\% &  6.0\% & 49.5\%&  4.8\% & 1.3\% &  5.4\% & 50.8\% \\
60  &  6.6\%  &  0.6\% &  7.4\% & 45.3\%&  4.3\% & 1.0\% &  5.0\% & 54.7\% \\
100 &  6.5\%  &  0.5\% &  7.5\% & 45.5\%&  5.3\% & 0.6\% &  6.1\% & 51.5\% \\
200 &  7.8\%  &  0.9\% &  8.8\% & 37.3\%&  6.2\% & 0.9\% &  7.2\% & 44.2\% \\
400 &  7.1\%  &  0.8\% &  7.5\% & 43.5\%&  7.2\% & 0.7\% &  8.1\% & 41.5\% \\
600 &  -  & - &  - & - &  6.7\% & 0.5\% &  7.0\% & 50.8\% 
\end{tabular}
\caption{Errors obtained from averaging 100 runs of the comparator, using
  direct and linear encoding after $10^7$ steps, for different input sizes $N$.
  The connection probabilities used are $p_{conn}^{(1)}=0.3$,
  $p_{conn}^{(2)}=0.8$. For $N>5$ the standard deviations amounts to
  0.1-0.8\% (decreasing with $N$) for the errors E, FP and FN, and 1\% for MI\%.
  For the $N=5$ case, the standard deviation of the errors is
  5-14\% (again, decreasing with $N$) while for MI\% it amounts to 15\%.}
\label{tab:errs}
\end{table}

We found that the performance of the comparator 
depends substantially on the degree of similarity 
of the two input signals $\mathbf{y}$ and $\mathbf{z}$ 
for the case when the two inputs are uncorrelated. For
a quantitative evaluation of this dependency we define the
Euclidean distance
\begin{equation}
d = \left|\left|\mathbf{y}-\mathbf{f}^{-1}(\mathbf{z})\right|\right|\;,
\label{eq:d}
\end{equation}
where $||\cdot||$ denotes the Euclidean norm of a vector.
For small input sizes $N$, a substantial fraction
of the input vectors are relatively similar with small Euclidean
distance $d$, resulting in a small output $x^{(4)}$.
This can prevent the comparator from learning the classification
effectively, thus the best accuracy is obtained for input vectors 
of size greater than $N=10$, compare table~\ref{tab:errs}. 

\begin{figure}[t] \centering
\includegraphics[width=0.9\textwidth]{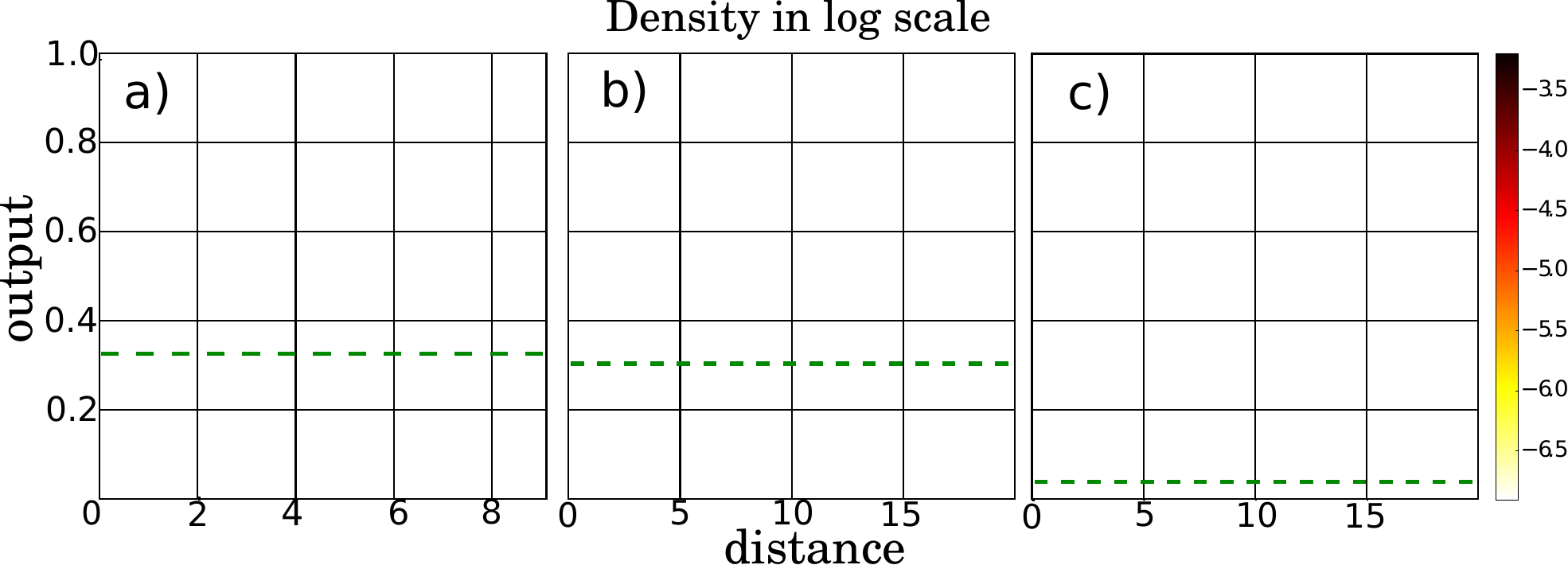}
\caption{The probability of an output $x^{(4)}$ (\ref{eq:max})
to occur as a function of Euclidean distance $d$ 
(\ref{eq:d}) of the input pairs $\mathbf{y}$ and $\mathbf{z}$
(see fig.~\ref{fig:scheme}), as a density plot (color coded),
using $N=400$, $\alpha=1$ and direct encoding. The last 
10\% of simulations with $10^7$ input pairs have been 
taken for the performance testing.
{\bf a)} Unconstrained input protocol (\ref{eq:random:unconstrained}) both
   for training and for testing
{\bf b)} Unconstrained input protocol (\ref{eq:random:unconstrained}) for
     training and constrained (\ref{eq:random:constrained}) for testing.
{\bf c)} Constrained input protocol (\ref{eq:random:constrained}) for both
   training and testing.
} 
\label{fig:bigsize} 
\end{figure}

The above phenomenon can be investigated systematically by
considering two distinct distributions for the
Euclidean distance $d$. Within our input protocol
(\ref{eq:z}) the pairs $\mathbf{y}$ and $\mathbf{z}$ 
are statistically independent with probability 
$(1-p_{eq})$. We have considered two ways of generating
statistically unrelated input pairs,
\begin{equation}
\parbox{0.7\textwidth}{{\em Unconstrained:}
      The components $y_i$ and $z_i$ are selected
      randomly from the interval $[-1,1]$.}
\label{eq:random:unconstrained}
\end{equation}
and
\begin{equation}
\parbox{0.7\textwidth}{{\em Constrained:}
      The components $y_i$ and $z_i$ are selected
      randomly such that the distance $d$ has a flat
      (uniform) distribution in $[0,1]$.}
\label{eq:random:constrained}
\end{equation}
For the case of the `unconstrained' input protocol the distribution
of distances $d$ is sharply peaked for large input size $N$,
compare fig.~\ref{fig:bigsize}. The impact of the
distribution of Euclidean distances between the random
input vectors $\mathbf{y}$ and $\mathbf{z}$
is presented in fig.~\ref{fig:bigsize}, where we show the
result of three separated simulations:
\begin{itemize}

\item[a)] Using the unconstrained input protocol
(\ref{eq:random:unconstrained}) 
     for both training and for testing. The corresponding performance
      errors 
     are $FP =1.0\%$, $FN =10.7\%$ and  $E =9.7\%$, for a 
     threshold $\theta=0.34$.
\item[b)] Using the unconstrained input protocol
(\ref{eq:random:unconstrained}) 
     for training and the constrained (\ref{eq:random:constrained}) protocol for 
     testing. The performance errors are $FP =13.7\%$, $FN =14.6\%$ 
     and $E =14.2\%$, for a threshold $\theta=0.28$.
\item[c)] Using the constrained input protocol
(\ref{eq:random:constrained}) 
          for both training and testing. The corresponding
          errors are $FP =79.9\%$, $FN =0.0\%$ and  $E =79.9\%$, 
          for a threshold $\theta=0.02$.
\end{itemize}
The accuracy of the comparator is very good for a). In this
case values close to $d \sim 0$ are almost inexistent 
for random input pairs $\mathbf{y}$ and $\mathbf{z}$,
random and related input pairs are clustered in distinct
parts of the phase space.

The performance of the comparator drops, on the other side, with
increasing number of similar random input pairs. For the case c) the
distribution of distances $d$ is uniform and the comparator has
essentially no comparison capabilities. Since the 20\% of the input is
correlated, the minimal error $E$ in this case is obtained if the system
assumes all input to be uncorrelated (i.e. setting an extremely small
threshold). That situation results in 80\% $FP$ and 20\% $FN$.
Notice that in this case the mutual information of the system is null.
Lastly, in the mixed case b) the comparator is trained with a
unconstrained distribution for the distances $d$ and tested using a
constrained distribution.  In this case the comparator still acquires
a reasonable accuracy of $E=14\%$.

\subsubsection{Equilibrated Input, $p_{eq}=0.5$}

In this subsection we expand the results for equilibrated input
data sets, viz $p_{eq}=0.5$ in (\ref{eq:z}). The procedure remains as
described in the previous section. Again, each calculation consists of
$t_{max}=10^7$ steps, from which the last 10\% of the simulation is
used for performance evaluation. This result is consistent with
the intuitive notion, that it is substantially harder to learn when
$\mathbf{y}$ and $\mathbf{z}$ are related, when most of the input 
stream is just random noise and semantically correlated input pairs 
occur only seldom. For applications one may hence consider a
training phase with a high frequency $p_{eq}$ of semantically 
correlated input pairs.

\begin{table}[t]\centering
\begin{tabular}{c||c|c|c|c}
$N$ & $\langle E \rangle$ & $\langle FP \rangle$ & $\langle FN
\rangle$ & MI\% \\\hline\hline
5   & 9$\pm$6 \%& 8$\pm$7 \%& 10$\pm$5 \%&  58$\pm$14 \% \\
15  & 3.9$\pm$0.4 \%& 0.4$\pm$0.1 \%& 6.9$\pm$0.6 \%& 78$\pm$1 \% \\
30  & 3.4$\pm$0.2 \%& 0.3$\pm$0.1 \%& 6.3$\pm$0.3 \%& 81$\pm$1 \% \\
60  & 3.3$\pm$0.1 \%& 0.2$\pm$0.1 \%& 6.1$\pm$0.2 \%& 81$\pm$1 \% \\
100 & 3.4$\pm$0.1 \%& 0.2$\pm$0.1 \%& 6.2$\pm$0.1 \%& 82$\pm$1 \% \\
200 & 4.7$\pm$0.1 \%& 0.5$\pm$0.3 \%& 8.2$\pm$0.5 \%& 75$\pm$1 \% \\
400 & 6.2$\pm$0.1 \%& 0.4$\pm$0.1 \%& 10.9$\pm$0.1 \%& 70$\pm$1 \% \\
600 & 7.5$\pm$0.1 \%& 1.1$\pm$0.1 \%& 12.4$\pm$0.1 \%& 66$\pm$1 \% 
\end{tabular}
\caption{Errors obtained from averaging 100 runs of the comparator, using
		linear encoding, with $p_{eq}=0.5$, for different input sizes $N$.
		The connection probabilities used are $p_{conn}^{(1)}=0.3$,
		$p_{conn}^{(2)}=0.8$. }
\label{tab:errs50}
\end{table}

As seen in table~\ref{tab:errs50}, the use of a balanced input set
does not change the general behavior but results in a substantial
increase in performance. The accuracy of the system in terms of
percentage of correct classifications (above 95\% accuracy except on
very small input size) and relative mutual information MI\% ($\sim
80\%$ of the maximum information gain) is very high. A relative mutual
information of MI\%$\approx$80\% means that the system recovers over
92\% of the maximally achievable correlations between the input and
the output, as shown in fig.~\ref{fig:MI_relative}.


\subsection{Effect of noisy encoding}

In the previous sections we have provided results showing that the
proposed comparator can achieve a good accuracy despite the fact that
a large part of the input is noise. In addition, the comparator is
also robust against a level of noise in the encoding of the inputs.
Random noise in the encoding would correspond to the neural
populations having rapid random reconfigurations or random changes in
the individual neurons' behavior above a certain level.

\begin{figure}[!ht]
\centering
\includegraphics[width=0.5\textwidth]{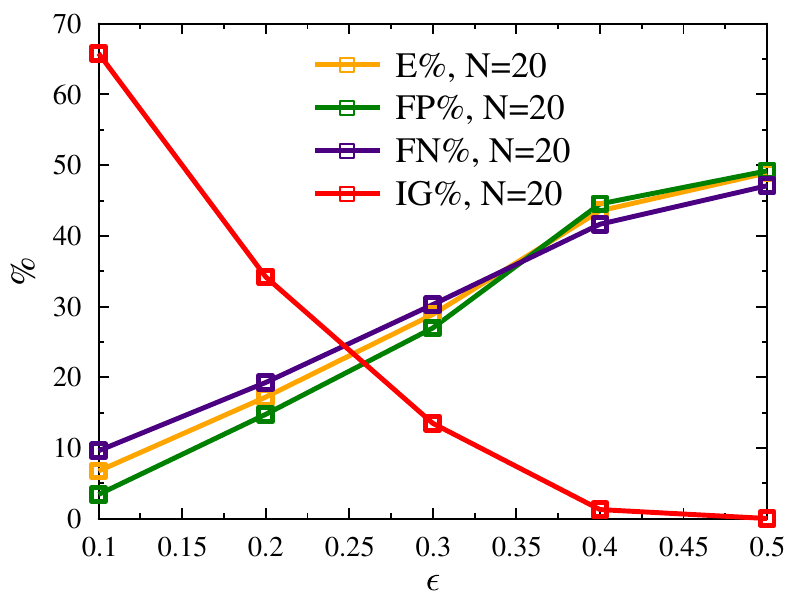}
\caption{Errors $E$, $FP$, $FN$, and percentage of information
gain MI\% for different levels of noise $\epsilon$ in the encoding.
$\epsilon=0.5$ corresponds to a magnitude equal to the average
activity of an input.}
\label{fig:noisy}
\end{figure}

As shown in fig.~\ref{fig:noisy}, the system has an accuracy decay if
the encoding is affected by random noise of the same magnitude as the
average input activity  (0.5). For this calculation, we define the
random noise in the encoding as adding a random number between 0 and
$\epsilon$ to each element of one of the compared inputs, i.e. $y_i
\to y_i + r_i$, where $r_i \in [0,\epsilon]$. The values $r_i$ are
changed in every step of the calculation.

The addition of random noise in the encoding is effectively seen by
the system as a slightly different input. Since the system is designed
to classify inputs either into different or equal, a large level of
noise drives the system into classifying the input as different.
However, if the input is only slightly changed, the correlation is
still found by the comparator and the output remains under the
threshold for classification.

\subsection{Impact of the frequency of correlated input and input size}
\label{sec:impactpeqandn}

\begin{figure}[!ht]
\centering
\includegraphics[width=0.49\textwidth]{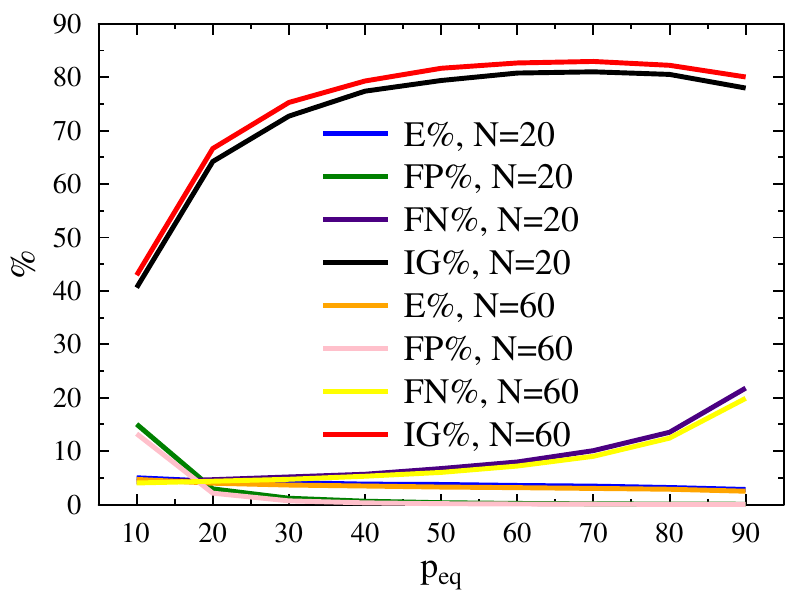}
\includegraphics[width=0.49\textwidth]{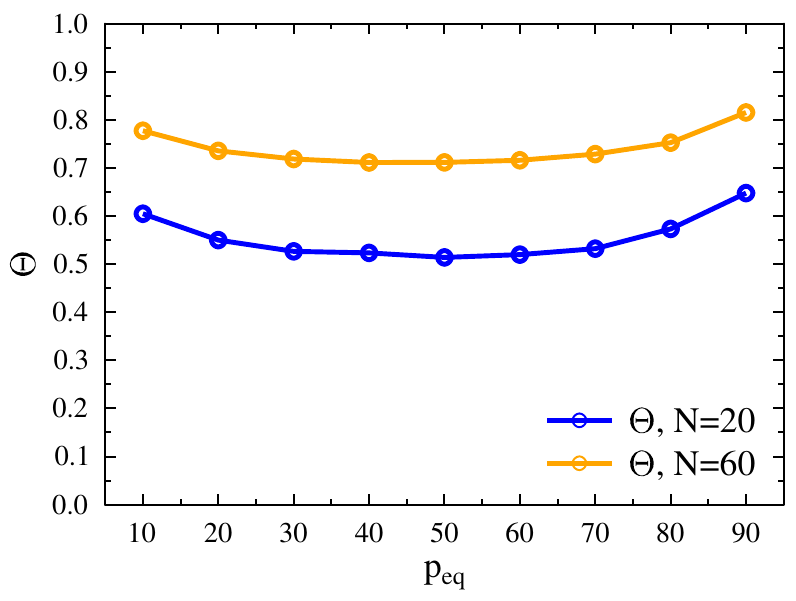}
\caption{On the left, the errors $E$, $FP$, $FN$, and relative
			mutual information MI\% for different frequencies of correlated input
			$p_{eq}$ (percentages). The results are presented for input sizes $N=$
			20 and 60. On the right, the optimal values of the threshold $\theta$
			for  different frequencies of correlated input $p_{eq}$. The results
			are again presented for input sizes $N=$ 20 and 60.}
\label{fig:peq}
\end{figure}

In fig.~\ref{fig:peq}, the dependency of the optimal threshold
$\theta$ and the errors $E, FP, FN,$ MI\% with the probability $p_{eq}$
is shown.  At a constant input size, the threshold shows only a weak
dependence with the probability $p_{eq}$. The threshold changes at its
maximum for the probability of any case in the order of 10\% or less.
The threshold varies less than 0.1 from $p_{eq}=0.1$ to $p_{eq}=0.9$.
This indicates that the system would still be effective if the
probabilities of the events change significantly, even without
readjusting the parameters $\alpha$ or $\theta$, or with a small
readjustment if the change is extreme.

\begin{figure}[!b]
\centering
\includegraphics[width=0.49\textwidth]{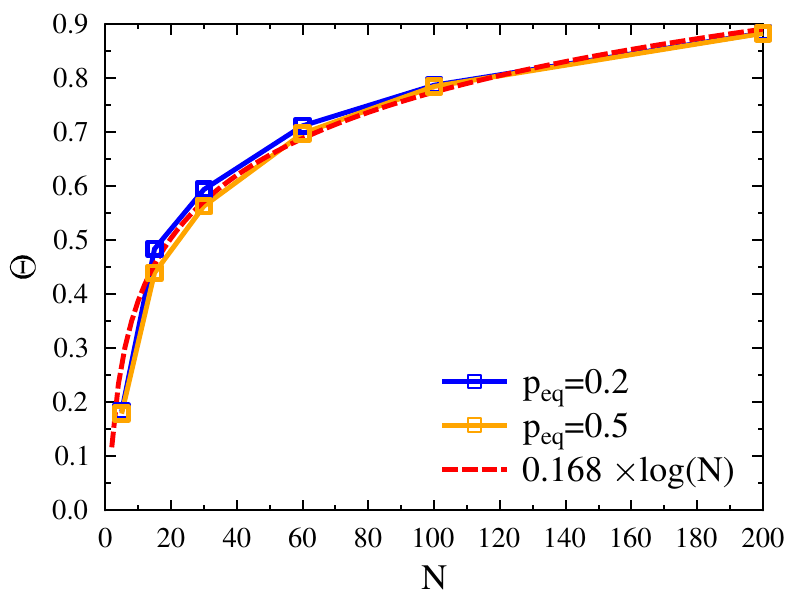}
\caption{Optimal threshold $\theta$ vs input size $N$, for
		$\alpha=2.7$. The behavior is markedly logarithmic.}
\label{fig:thetaVsN}
\end{figure}

In fig.~\ref{fig:thetaVsN}, the dependence of the optimal threshold
$\theta$ with the input size $N$ is presented. The threshold has a
marked logarithmic dependency with respect to the system size. In
effect, the threshold $\theta$, the gain $\alpha$ and the system size
$N$ are all strongly coupled, such that given an input size the rest
of the parameters are essentially fixed. 

\subsection{Comparison of inputs with different sizes}

\begin{table}[!hb]
\centering
\begin{tabular}{c||c|c|c|c||c|c|c|c}
 & \multicolumn{4}{|c||}{$N=20$} & \multicolumn{4}{|c}{$N=60$} \\
 $\Delta$ & $\langle E \rangle$ & $\langle FP \rangle$ & $\langle FN
\rangle$ & MI\% & $\langle E \rangle$ & $\langle FP \rangle$ &
$\langle FN \rangle$ & MI\% \\\hline\hline
 0 & 3.4$\pm$0.3 & 0.3$\pm$0.1 & 6.3$\pm$0.4 & 80$\pm$1  &  3.3$\pm$0.1 & 0.2$\pm$0.1 & 6.1$\pm$0.2 & 81$\pm$1 \\
 5 & 3.1$\pm$0.3 & 0.3$\pm$0.1 & 5.7$\pm$0.5 & 82$\pm$1  &  3.4$\pm$0.3 & 0.3$\pm$0.1 & 6.3$\pm$0.2 & 80$\pm$1 \\
10 & 2.7$\pm$0.2 & 0.3$\pm$0.1 & 4.9$\pm$0.4 & 84$\pm$1  &  2.9$\pm$0.1 & 0.2$\pm$0.1 & 5.4$\pm$0.2 & 83$\pm$1 \\
20 & 2.2$\pm$0.2 & 0.3$\pm$0.1 & 4.0$\pm$0.4 & 86$\pm$1  &  2.6$\pm$0.1 & 0.2$\pm$0.1 & 4.8$\pm$0.1 & 85$\pm$1 \\
40 & 1.6$\pm$0.1 & 0.3$\pm$0.1 & 2.9$\pm$0.3 & 89$\pm$1  &  2.2$\pm$0.1 & 0.2$\pm$0.1 & 4.0$\pm$0.1 & 87$\pm$1
\end{tabular}
\caption{Average errors  $\langle E \rangle$, $\langle FP \rangle$, $\langle
		FN \rangle$ and relative mutual information MI\% obtained from 100
		runs of the comparator for comparing one input of size $N=20$ ($N=60$, right)
		and another of size $N+\Delta$, using $p_{eq}=50$.}
\label{tab:diffsizes}
\end{table}

The comparator successfully compares input of different sizes. In
table~\ref{tab:diffsizes} we show the average accuracy over 100 runs
of a comparator where one of the vectors to be compared has a size
$N$ and the other has a larger size $N+\Delta$. The number of extra
inputs $\Delta$ is maintained constant during the whole simulation. In
each step, the values of the two vectors are assigned as described
previously as ``linear encoding'' in sec.~\ref{sec:protocol}. The
linear encoding is done in this case with a matrix $\hat{A}$ that has
dimensions $(N+\Delta) \times N$, thus the information gets encoded in
a vector of higher dimension.

The accuracy of the comparator does not decrease, but, rather
surprisingly, it slightly increases. There is no loss in accuracy
because the uncorrelated inputs are not minimized to a value close to
zero due to the anti-Hebbian adjustment of the synaptic weights, as
happens only with the correlated input. We attribute the small
increase in accuracy to the increase of neurons involved in the
system.

\subsection{Influence of connection density}

A key ingredient in this model is the suppression of a 
fraction of inter-layer connections with probability $1-p_{conn}$,
which is necessary to give higher-layer neurons the
possibility to encode varying features of correlated
input pairs. For a systematic study we ran
simulations using a range of distinct probabilities
of interconnecting the layers.

\begin{figure}[t]
\centering
\includegraphics[width=\textwidth]{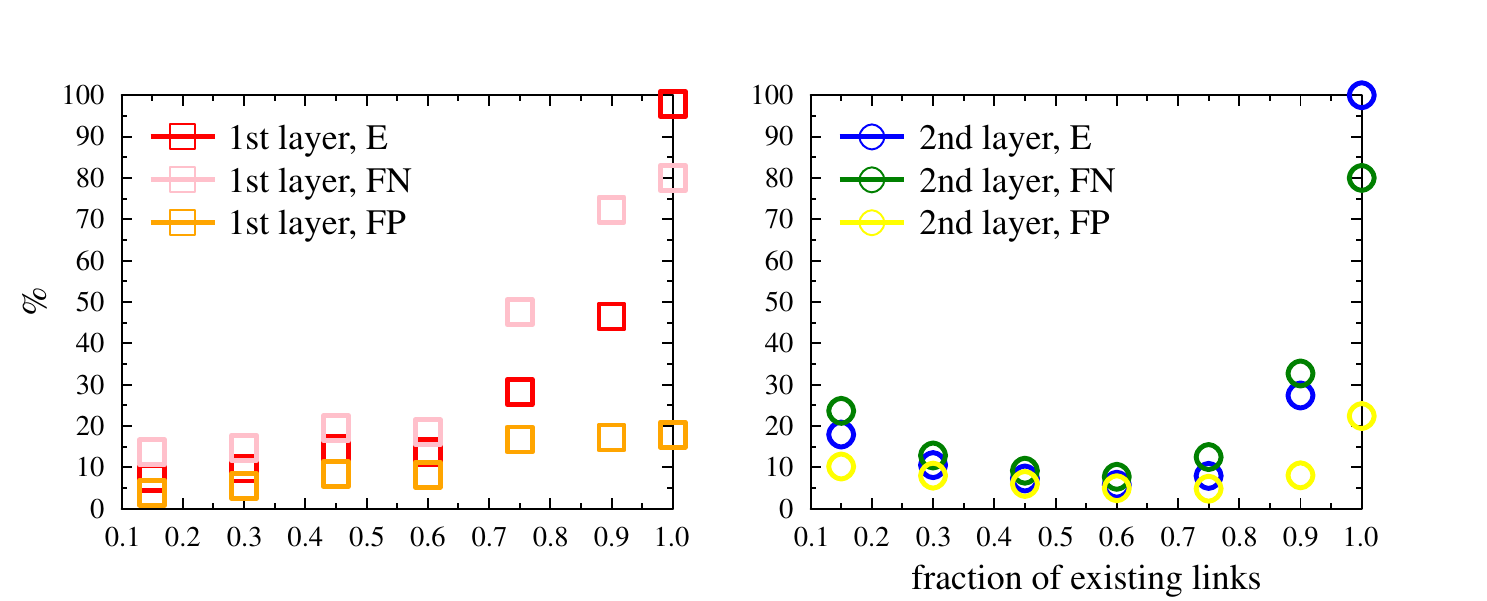}
\caption{Effect of the probability of connection in the overall
errors. On the left, the errors of the comparator with varying
probability $p_{conn}^{(1)}$ of connection efferent 
from the first layer, with $p_{conn}^{(2)}=0.75$. 
On the right, the probability $p_{conn}^{(2)}$ of 
efferent connection from the second layer is varied,
keeping $p_{conn}^{(1)}=0.3$ constant.
         }
\label{fig:probconnavgs}
\end{figure}

In figure~\ref{fig:probconnavgs}, we show the unconstrained
performance measures for $N=5$ when
changing (left) the connection $p_{conn}^{(1)}$
from the input layer to the first layer (compare
fig.~\ref{fig:scheme}, with constant
$p_{conn}^{(2)}=0.75$) and (right) when
varying the connection $p_{conn}^{(2)}$
from the second to the third layer. In the later
case we kept $p_{conn}^{(1)}=0.3$ fixed.

The data presented in fig.~\ref{fig:probconnavgs}
show that the neural comparator loses functionality
when the network becomes fully interconnected. The optimal 
interconnection density varies from layer to layer
and is best for 10\% efferent first-layer connections
and 60\% links efferent from the second layer.

\subsection{Images Comparison}

\begin{figure}[t]
\centering
\includegraphics[width=0.8\textwidth]{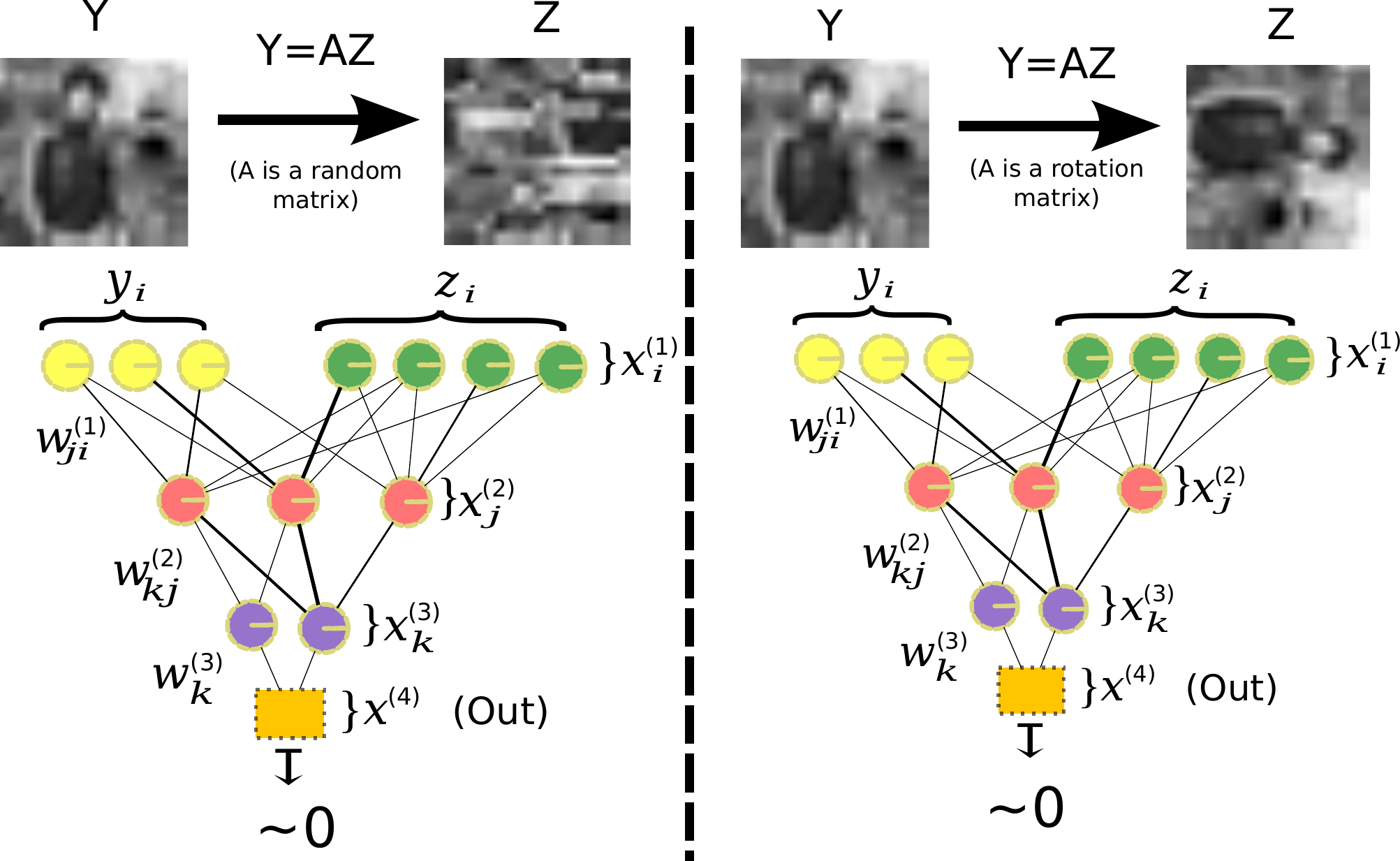}
\caption{Scheme of two possible encryptions made by the comparator
with probability $p_{eq}$. In both cases, after being exposed to many
images the comparator would ideally result in an output $x^{4} \sim
0$. Notice the matrix $A$ in each side of the figure is constant
during the whole simulation.}
\label{fig:imgcomp}
\end{figure}

We tested the comparator efficiency in comparing a set of black and
white pictures of small size (20$\times$20 pixels, i.e. $N=400$) using linear
encoding via a random matrix as in previous sections, see
fig.~\ref{fig:imgcomp}. The set of pictures is very small (200
pictures) in comparison to the input data used to train the comparator
($t=10^7$ inputs). The results can be seen in table~\ref{tab:img}. The
limited input set has the negative effect that the comparator is not
able to learn comparison only from this set. This suggests that in
order for the comparator to develop its functionality, it must
sample a sizable part of the possible input patterns.

\begin{table}[!ht]
\centering
\begin{tabular}{c||c|c|c|c}
  & $\langle E \rangle$ & $\langle FP \rangle$ & $\langle FN
\rangle$ & MI\% \\\hline\hline
  Only Images $p_{eq}=0.2$ & 20.6$\pm$0.3 & 51.9$\pm$0.6 & 14.4$\pm$0.2 & 8$\pm$1 \\\hline
  Trained w/Random input $p_{eq}=0.2$ & 14.5$\pm$0.2 & 39.3$\pm$0.3 & 6.2$\pm$0.1 & 32$\pm$1 \\\hline
  Trained w/Random input $p_{eq}=0.5$ & 10.8$\pm$0.1 & 10.7$\pm$0.2 & 10.9$\pm$0.1 & 51$\pm$1 \\\hline
\end{tabular}
\caption{Average errors  $\langle E \rangle$, $\langle FP \rangle$, $\langle
FN \rangle$ and relative mutual information MI\% obtained from 100
runs of the comparator for comparing black and white pictures of size
20$\times$20 pixels.}
\label{tab:img}
\end{table}

As explained previously, the correlated inputs are minimized 
by the anti-Hebbian rule, while the uncorrelated input cannot be 
minimized to the same level, since those cases result in the terms 
$x_i^{(k)}x_j^{(k+1)}$ in eq.~(\ref{eq:updatew}) being essentially 
random. This assumption is however not fulfilled if the values of 
these terms are not well distributed (unless their values are by 
chance always small), which is the case if the sampling is not 
large enough.

As a second test, we initially trained the comparator using random
data (still using $p_{eq}=0.5$) in order to start with a functional
distribution of the synaptic weights, and then switched to the picture
set for the last 10\% of the calculation, with the comparator still
learning during this stage. In this case, the comparator achieved its
function (see table~\ref{tab:img}). However, the accuracy did not
fully reach that of the system when comparing randomly generated data.

We expect the accuracy of the random comparator to be at the level of
the generated by random input if the input stream explores a sizable
part of the possible input. For instance, ideally the image input
would be a video of the visual input in a mobile agent while exploring
the environment, such that a large amount of patterns are processed by
the comparator. This is however out of the scope for this work, while
follow up work is expected.

\section{Interpretation within the scope of fuzzy logic}
\label{sec:fuzzy}

\begin{figure}[t]
\centering
\includegraphics[width=0.9\textwidth]{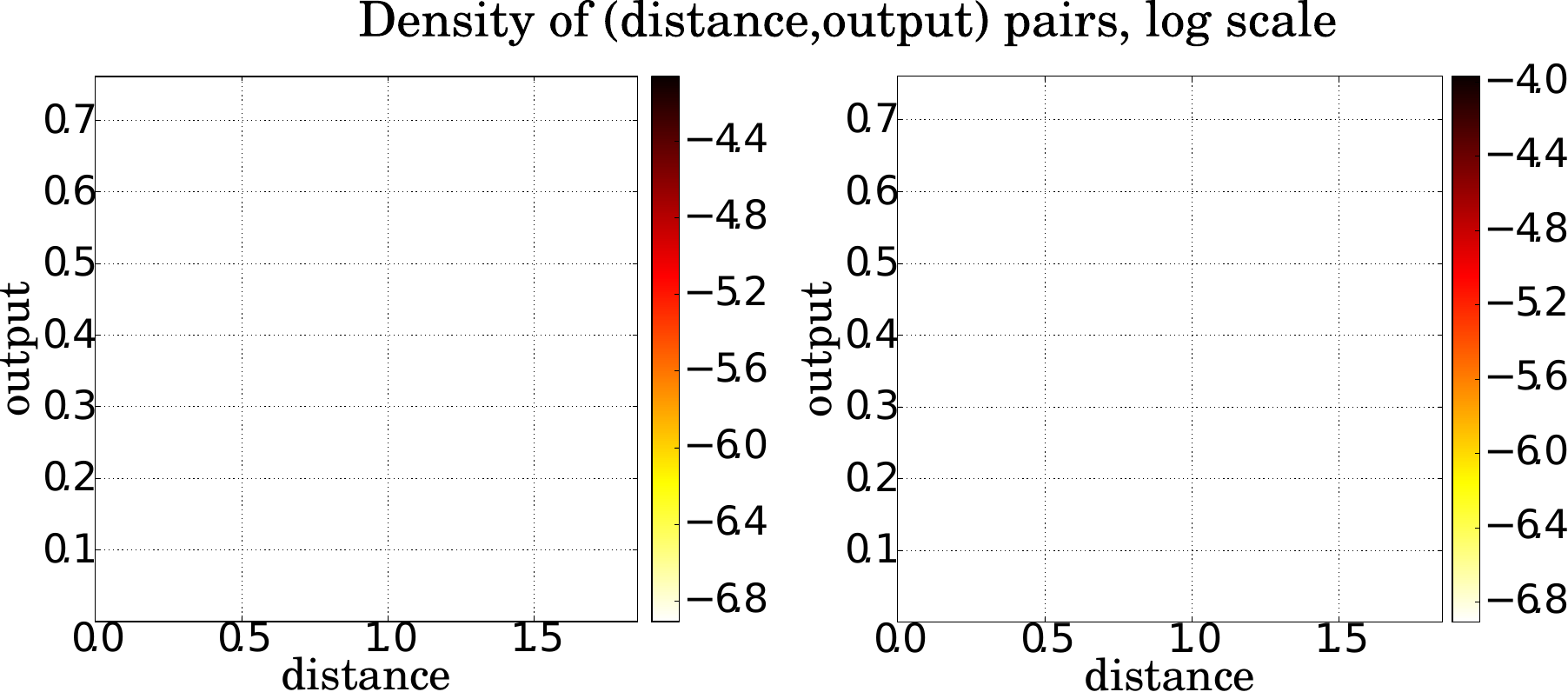}
\caption{Density of output-distance pairs from two 
individual runs of the comparator, comparing inputs 
of dimension $N=5$ under the direct encoding (left) 
and linear encoding (right), using $\alpha=2.7$. 
}
\label{fig:evendens}
\end{figure}

The dependency of the output of the comparator seen in
fig.~\ref{fig:bigsize}b,c and fig.~\ref{fig:evendens} can be
interpreted in terms of fuzzy logic \citep[][]{Keller1992}, offering
alternative application scenarios for the neural comparator.

The error measures evaluated in table \ref{tab:errs}, like the
incidence of false positives ($FP$), are based on boolean logic, the
classification is either correct or incorrect, i.e. binary. For
real-world applications the input pairs $(\mathbf{y},\mathbf{z})$ may
be similar but not equal and the dependency of the output as a
function of input similarity is an important relation characterizing
the functionality of neural comparators.

The comparator essentially provides a continuous variable classifying
how much the input case corresponds to the case of equal input, i.e. a
truth degree. Thus, the comparator can be interpreted as a fuzzy logic
gate for the operator ``equals'' ($=$), since it provides a truth
degree for the outcome of the discrete version of the same operator.

In fig.~\ref{fig:evendens} we present, on a logarithmic scale, the
density of results for the observed output $x^{(4)}$, as a function of
the distance $d$ between the respective inputs, for one single run of
the comparator. $80\%$ of the input vectors were randomly drawn and
later readjusted in order to fill the range of distances $d=0.1$ to
$1.5$ uniformly, according to the constrained protocol
(\ref{eq:random:constrained}).  In addition, $20\%$ of the input have
a distance of $d=0$ with $\mathbf{z}=\mathbf{y}$, resulting in the
high density of simulations at $d=0$.
 
\begin{figure}[t]
\centering
\includegraphics[width=0.6\textwidth]{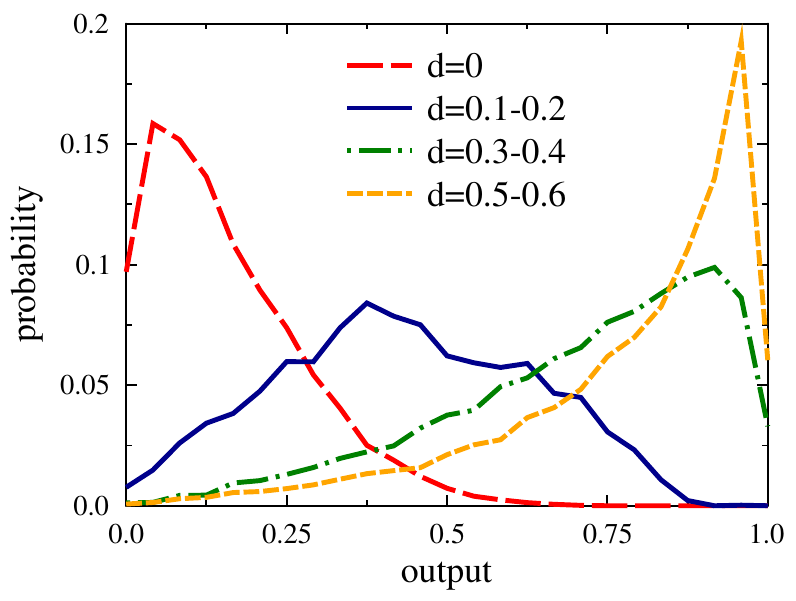}
\caption{Output distribution for different Euclidean 
         distances $d$ (\ref{eq:d}) between the inputs, 
         as a function of output, corresponding to a
         vertical slice of the $N=5$ direct encoding  
         data presented in fig.~\ref{fig:evendens}.}
\label{fig:distr-out-d}
\end{figure}

The uncertainty of the classification of inputs presented in
fig.~\ref{fig:evendens} is reflected in a probability distribution for
the comparator output, shown in fig.~\ref{fig:distr-out-d} for the
case of direct encoding. The output distribution is narrower for cases
where the distance $d$ corresponds to clearly correlated or
uncorrelated inputs.

The distributions presented in fig.~\ref{fig:evendens} can be
interpreted as fuzzy probability distributions for any given distance
$d$ (vertical slices), as shown in fig.~\ref{fig:distr-out-d}.  The
probability for the input pairs $\mathbf{y}$ and $\mathbf{z}$ to be
classified as different decreases with decreasing distance $d$ between
them. This shows that inputs with smaller distances have in general
increasingly weaker outputs. Thus, assuming that the Euclidean
distance $d$ is a good estimator of how similar the input is, the
output of the comparator provides an arguably reliable continuous
variable estimating a similarity degree for the inputs, i.e.  the
truth degree of the operator ``equals'' applied to the inputs.

\section{Discussion}

The results presented here demonstrate that the proposed 
neural comparator has the capability of discerning similar 
input vectors from dissimilar ones, even under noisy
conditions. Using 80\% noise, with four out of five inputs being randomly
drawn, the unsupervised comparator architecture achieves a boolean
discrimination accuracy of above $\sim$90\%.
The comparator circuit can also achieve the same accuracy when the
inputs to be compared are encoded differently. If the encodings of both
inputs are related by a linear relation, the accuracy of the
comparison does not worsen with respect to the direct encoding case.

A key factor for the accuracy of the method is the inclusion of a
slightly different path for the layer-to-layer information, provided
by random suppressions of interlayer connections.  However, the
suppression has the potential side effect of rendering some of the
correlations difficult to be learned. For this reason a compromise
needs to be found between the number of connections that must be kept in
order to maintain the network functional and the number of connections
that needs to be removed to generate sufficiently different outputs in
the third layer.

We find it remarkable that from a very simple model of interacting
neurons under the rule of minimization of its output, the fairly
complex task of identifying the similarity between unrelated inputs can
emerge through self-organization without the need of any predefined or
externally given information. Complexity arising from simple
interactions is a characteristic of natural systems, and we believe
the capacity of many living beings to perform comparison operations
could potentially be based on some of the aspects included in our
model.

\section*{Conclusion}

We have presented a neuronal circuit based on a feed-forward
artificial neural network, which is able to discriminate whether two
inputs are equal or different with high accuracy even under noisy
input conditions.

Our model is an example of how algorithmic functionalities can emerge
from the interaction of individual neurons under strictly local rules,
in our case the minimization of the output, without hard-wired
encoding of the algorithm, without external supervision and without
any \emph{a priori} information about the objects to be compared.  Since our
model is capable of comparing information in different encodings, it
would be a suitable model of how seemingly unrelated information
coming from different areas of a brain can be integrated and
compared.

We view the architecture proposed here as a first step towards an
in-depth study of the important question: which are possible
neural circuits for the unsupervised comparison of
unknown objects. Our results show, that anti-Hebbian adaption
rules, which are optimal for synaptic information transmission
\citep[][]{bell95}, allow to compare two novel objects, viz
objects never encountered before during training, with respect
to their similarity. The model is capable not only to provide binary
answers -- whether the two objects in the sensory stream are (are not)
identical -- but also to give a quantitative estimate of the
degree of similarity, which may be interpreted in the context
of fuzzy logic. We believe this quantitative estimate
of similarity to be a central aspect of any neural comparator,
as it may be used as a learning or reenforcement signal.

\section*{Acknowledgments}
We would like to acknowledge the support of the German Science
Foundation.

\end{document}